\documentstyle[aps,preprint,epsfig]{revtex} 
\tightenlines
\begin{document}

\newcommand{\be}{\begin{equation}}
\newcommand{\ee}{\end{equation}}
\newcommand{\ket}[1]{\left|#1\right>}
\newcommand{\bra}[1]{\left<#1\right|}

\bibliographystyle{unsrt}
\title{Realism or Locality: Which Should We Abandon?} 
\author{Raymond Y. 
Chiao\footnote{On appointment as a Miller Research Professor in the Miller 
Institute for Basic Research in Science}
 and John C. Garrison\\ Department of Physics, University of 
California, Berkeley, California 94720-7300\\ Tel: (510) 642-4965, (510) 
642-5620, Fax: (510) 643-8497, E-mail: chiao@physics.berkeley.edu, 
garrison@physics.berkeley.edu} 
\date{August 15, 1998}
\maketitle
\begin{abstract}

We reconsider the consequences of the observed violations of Bell's 
inequalities. Two common responses to these violations are: (i) the 
rejection of realism and the retention of locality, and (ii) the rejection 
of locality and the retention of realism. Here we critique response (i). 
We argue that locality contains an implicit form of realism, since in a 
world view that embraces locality, spacetime, 
with its usual, fixed topology, 
has properties independent of 
measurement. Hence we argue that response (i) is incomplete, in that its 
rejection of realism is only partial.

\end{abstract}
\pagebreak

\section{Introduction}

It is a great pleasure to dedicate this paper to Prof. Daniel Greenberger 
on the occasion of this Festschrift in his honor. 

The proof of Bell's theorem \cite{Bell} is based on the two main 
assumptions of realism and locality. Theories which satisfy these two 
assumptions are commonly called ``local realistic theories.'' 

{\em Realism}, roughly speaking, is the belief that there exists an 
objective world ``out there'' independent of our observations. This is the 
world view of classical physics. In an extension to quantum physics, this 
world view was expressed in the classic paper of Einstein, Podolsky and 
Rosen (EPR), when they made the following statement~\cite{EPR}:

\begin{quote}
``If, without in any way disturbing a system, we can predict with 
certainty (i.e., with probability equal to unity) the value of a physical 
quantity, then there exists an element of physical reality corresponding 
to this physical quantity.'' \end{quote} 

\noindent In this formulation the authors have identified a class of 
situations for which, in their view, it would be absurd to deny 
independent physical reality~\cite{d'Espagnat} to a property of the 
system. This is not to say that there are no other ``elements of physical 
reality'' with definite (but possibly unknown and unpredictable) values 
which do not depend on measurement. We shall use the term {\em elements of 
physical reality} in this more general sense, reserving the term {\em EPR 
elements of physical reality} for the specific features described in the 
quotation. Thus in a realist's world view, there exist physical quantities 
with objective properties, which are independent of any acts of 
observation or measurement. For example, consider the case of two spin 1/2 
particles emitted in opposite directions from a common spin zero source 
prepared in the Bohm singlet state~\cite{Bohm},

\be
\ket{S_{total}=0} =
\frac{1}{\sqrt{2}}\{\ket{S_{z1}=+1/2}\ket{S_{z2}=-1/2} - 
\ket{S_{z1}=-1/2}\ket{S_{z2}=+1/2}\}.
\ee

\noindent When a measurement of one particle, by a detector located at the 
end of one arm of the experimental apparatus, yields a definite result of 
spin $S_{z}=+1/2$, then conservation of the total spin angular momentum of 
the system necessarily requires that immediately upon this measurement, 
the other particle must definitely possess a spin $S_{z}=-1/2$, and that 
this result must be independent of whether or not any measurement is 
actually made on the other particle. According to EPR, this implies that 
the other particle must actually possess spin $S_{z}=-1/2$ as an EPR 
element of physical reality.

{\em Locality} is the other main assumption which goes into the proof of 
Bell's theorem.  
the 
The idea here is that what happens in a 
finite spacetime region $A$ cannot be affected by what is done (e.g., 
by the 
choice of settings of measuring devices) in a spatially well-separated 
spacetime region $B$.  
In the context of special relativity, ``spatially 
well-separated'' means that all points of $A$ have spacelike separations 
from all points of $B$. Thus the locality principle is usually assumed to 
be equivalent to {\em relativistic causality}, i.e.,
the nonexistence of controllable superluminal signals.  
Combining the idea of realism 
with the locality principle leads to a strong separability condition, 
i.e., the factorizability of joint probabilities of spacelike separated 
coincidence detections. This factorizability is the most important 
technical ingredient in the proof of Bell's theorem.

Many experiments have shown that the Bell's inequalities, which follow 
from Bell's theorem, are violated (apart from technical caveats concerning 
detection loopholes, etc.), and that the predictions of quantum theory, 
which also violate these inequalities, are confirmed. These observations 
imply that some or all of the assumptions used in the proof of Bell's 
theorem are inconsistent with experiment. In addition to realism and 
locality, there are auxiliary (but not necessarily independent) 
assumptions, which include absence of advanced actions, contrafactual 
definiteness, etc. In accord with the usual practice in this discussion, 
we shall accept the
auxiliary
assumptions and
concentrate on realism and locality. With this understanding the following 
alternatives exist:

(i) Reject realism and retain locality (the {\em localist} position). 

(ii) Reject locality and retain realism (the {\em realist} position). 

(iii) Reject both realism and locality (the {\em nihilist} position). 

\noindent The nihilist position is the most difficult to analyze, since it 
contains no positive statement of what is to be retained. Thus an 
application of Occam's razor suggests that the localist and realist 
options should be examined first. 

\section{The localist position}

Here we would like to critique the localist position, 
which consists of two components.  The first is a 
denial of the existence of EPR elements of reality, and hence by 
implication, of realism in general.  The second is 
an affirmation of relativistic causality, as embodied in special 
relativity and the standard notions of causality,
and hence of the locality principle.  
Einstein foresaw 
this position in his 1948 letter to Born as follows~\cite{Einstein}:

\begin{quote}
``There seems to me no doubt that those physicists who regard the 
descriptive methods of quantum mechanics as definitive in principle 
would react 
to this [EPR] line of thought 
in the following way: they would 
drop the requirement for the independent existence of the physical 
reality present in different parts of space; they would be justified 
in pointing out that the quantum theory nowhere makes explicit use of 
this requirement.''
\end{quote}

\noindent In this quote, Einstein predicted that those physicists who
believe 
that quantum theory is fundamental would abandon (``drop'') realism 
(``the independent existence of the physical reality present in 
different parts of space'').  However, these same physicists would of 
course not abandon special relativity, which has been so amply 
validated by experiment.  Thus Einstein presciently described the 
localist response to Bell's theorem.  

In his review article~\cite{Mermin}, ``Is the moon there when nobody 
looks?  Reality and the quantum theory,'' written after Bell's 
discovery, Mermin described the situation as follows:

\begin{quote}
``Using a {\em gedanken} experiment invented by David Bohm, in which 
`properties one cannot know anything about' (the simultaneous values 
of the spin of a particle along several distinct directions) are 
required to exist by the EPR line of reasoning, Bell showed (`Bell's 
theorem') that the {\em nonexistence} [emphasis added] of these 
properties is a direct consequence of the quantitative numerical 
predictions of quantum theory." 
\end{quote}

\noindent This rejection of the kinds of realistic elements required 
by the EPR line of reasoning emphasizes, at least on the surface, the 
apparent impossibility of a realistic account of microscopic 
phenomena.  
The combination of this anti-realistic line of reasoning 
with the evident 
importance of relativistic causality explains the strong attraction 
that the localist position holds for many physicists.

We argue here that this position is incomplete in its rejection of 
realism, since the concept of locality, as applied to EPR experiments, 
must implicitly carry within it a realist's view of space and time.

\section{Spacetime as an element of physical reality} 

Implicitly in the localist's world view, the spacetime manifold itself is 
an important element of physical reality in the EPR experiments.
In all currently successful
physical theories, e.g., the Standard Model of particle physics, spacetime 
is treated as if all its properties are independent of observation or 
measurement; hence it is indeed an element of physical reality in the 
general sense used here. In particular, the position of a photon detector 
in an EPR experiment is treated as if it were a classical quantity whose 
value does not depend on observations. Also, the time when a ``click'' 
goes off at this detector is a definite, realistic quantity. Thus an event 
defined by a particle detector firing at a given position and at a given 
time is assumed to be immune from any quantum back-actions, and is 
therefore viewed as constituting an element of physical reality. (Here the 
small back-actions of the detector on the particle~\cite{Aharonov} are 
neglected). 

In special relativity, the metric of the spacetime through which a photon 
propagates is strictly Minkowskian. Thus the Minkowskian (or flat) 
spacetime metric is also an element of physical reality which is immune 
from quantum back-actions arising from any acts of measurement. In 
particular, the causal light-cone structure of spacetime is a realistic 
concept which is implicitly assumed in the proof of Bell's theorem, in 
order to enforce the factorizability of the joint probability distribution 
of coincidence detections of EPR particles. This factorizability expresses 
the strong separability of spacelike-separated measurements. Also assumed 
is the absence of backwards-in-time causation (in other words, ``advanced 
actions''), i.e., the current values of the realistic variables do not 
depend on the values to be found in any future measurement. Thus in the 
proof of Bell's theorem, it is assumed from the start that spacetime, as 
embodied in special relativity and the standard notions of causality, is 
an element of physical reality. 

The localist position, in which realism is abandoned for material systems, 
but implicitly retained for the properties of spacetime, is not logically 
inconsistent. Indeed it may be viewed as a weaker version of the 
traditional Copenhagen interpretation in that spacetime is required to be 
described by a classical realistic theory, but all physical objects, 
including the measuring apparatus, would be described entirely by quantum 
theory. The real difficulty with this view is physical. We already know 
from classical general relativity that the spacetime manifold is itself 
dynamical. The question naturally arises: Should the spacetime manifold 
itself be treated by the quantum theory? This question is usually 
discussed in the context of quantizing the gravitational field, so that 
one would expect 
the EPR effects 
to occur only on Planck length scales rather 
than on the macroscopic length scales involved in EPR experiments. 
However, the correct answer may involve more radical, global alterations 
in the {\em topology} of spacetime (e.g., wormholes,
strings, etc.) 
\cite{Howard}, which 
have manifestations on macroscopic length scales. Thus the assumption of 
the standard fixed topology for spacetime used in both special and general 
relativity may no longer apply. Since quantum theory is believed to be 
universally applicable, the position we take here is that if one quantizes 
one part of system (matter), one should also {\em in principle} quantize 
the other part of system (spacetime), even if this is very difficult to 
accomplish in practice. 

The above approximations concerning the realistic nature of detector 
events and of the flat spacetime inside which they occur, are of course 
extremely good ones, and therefore are usually not explicitly stated. 
However, it is important to state them explicitly here, since these 
approximations carry with them a realist's world view of a very important 
part of the entire physical system, namely, the spacetime inside which the 
apparatus resides. To sum up: the localist position, in which realism is 
rejected for one part of the system, but retained for another part, 
represents {\em in principle} an incomplete point of view.

\section{The superposition principle implies superluminal phenomena}

Quantum theory 
predicts, and experiments demonstrate, that 
certain counterintuitive superluminal effects exist.
These effects arise from the tension between the {\em local} nature of all 
physical phenomena required by special relativity, and the {\em global} 
nature of superposed states required by quantum theory. However, these 
superluminal effects do not in fact violate relativistic causality due to 
the acausal nature of the correlations imposed by the superposition 
principle. The result is what Shimony has called ``peaceful coexistence'' 
between relativity and the quantum theory~\cite{Shimony}.

There exists evidence 
for quantum superluminal effects other than EPR, 
namely, in tunneling. Experiments have shown that individual photons 
penetrate a tunnel barrier with an effective group velocity which is 
considerably greater than $c$~\cite{tunneling1}. The experiments were 
conducted with a two-photon parametric down-conversion light source, which 
produced correlated, but random, emissions of photon pairs. The two 
photons of a given pair were emitted in slightly different directions so 
that one passed through a tunnel barrier, while the other passed through 
vacuum. The time delay for the tunneling photon relative to its twin was 
measured by adjusting the path length difference between the two photons 
in order to achieve coincidence detection. The photon transit time through 
the barrier was found to be smaller than the light transit time through an 
equal distance in the vacuum. However, according to standard theory there 
is no violation of relativistic causality, since the front velocity of the 
tunneling photon wave packet, which represents the speed at which an 
effect is connected to its cause, is still exactly $c$~\cite{tunneling2}.

Both the EPR and the tunnel effects represent examples of quantum 
superluminalities, but of different kinds. Tunneling involves one-particle 
states, while the EPR correlations involve many-particle states, so that 
the many-particle configuration space becomes important. The superposition 
principle of quantum theory is at the heart of both of these superluminal 
phenomena. In the case of the EPR effect, the superposition in Eq.~(1) 
leads to an ``entangled state,'' i.e., a superposition of product states, 
which results in nonfactorizable, i.e., nonseparable, joint probabilities, 
and thus superluminal correlations in coincidence detection. This {\em 
quantum nonseparability} is in direct contradiction to the strong 
separability, i.e., factorizability, assumption of Bell's theorem. In the 
case of the tunnel effect, superposition gives rise to exponential decay 
and spatial independence of the phase of the wave function inside the 
barrier. This leads to the superluminal behavior of the tunneling photon 
wave packet~\cite{tunneling1,tunneling2}.

\section{What is to be done?}

We have argued above that the rejection of realism in the localist 
position does not go far enough, since general relativity implies that 
spacetime itself must be treated as a dynamical quantum system. This 
argument seems to lead to a curious impasse, since denial of reality for 
spacetime makes it difficult to formulate the notion of locality which is 
such an important part of the localist position. Furthermore this approach 
seems to be perilously close to the nihilist position. If the localist 
position is to be maintained, it would seem to be necessary to formulate a 
notion of locality which does not require a realistic spacetime.  
Thus we may have to abandon the usual, fixed topology of spacetime. 

The realist position might offer an alternative in the form of an 
explicitly nonlocal theory which ascribes objective properties to material 
systems and spacetime alike. This approach has the obvious advantage that 
the existing notions of locality can still be formulated, but it poses the 
equally obvious difficulty of constructing a nonlocal theory which does 
not violate relativistic causality, at least in the approximation that 
spacetime fluctuations can be neglected. In particular this theory should 
satisfy the principle of ``peaceful coexistence'' with relativity, i.e., 
the explicitly superluminal (nonlocal) features of the theory should not 
permit the sending of {\em controllable} superluminal signals. It is 
difficult to see how this condition could be satisfied without taking the 
unpleasant step of postulating elements of physical reality which are {\em 
in principle} unobservable. It is an open question whether further 
development of the nonlocal, realistic quantum theory proposed by Bohm 
{\em et al.}~\cite{Bohm&Hiley} could lead to a model satisfying these 
constraints.

At the present time there does not seem to be any decisive argument for 
choosing between these alternatives. 

\section{Acknowledgments}

This work was supported by ONR grant no.~N000149610034 and by NSF 
grant no.~PHY-9722535.  R.~Y.~C. would like to thank the Miller 
Institute of Berkeley for support and the Universit{\'{e}} 
Interdisplinaire de Paris for an invitation to participate in the 1998 
International Colloquium, ``Aux Fronti{\`{e}}res de la Physique,'' and 
the participants of this colloquium for stimulating discussions.  We 
would also like to thank Bill Unruh, David Mermin, Henry Stapp, and 
Philippe Eberhard for their helpful comments on the manuscript.


\end{document}